%% file: paper_index.tex
\documentclass{article}
\usepackage{booktabs} % For formal tables
\usepackage{multicol}
\usepackage{multirow}
\usepackage[vlined,linesnumberedhidden,ruled,resetcount]{algorithm2e}
\usepackage{graphicx}
\usepackage{float}
\usepackage{xcolor}
\usepackage[labelsep=quad,indention=10pt]{subfig}
\usepackage[bottom]{footmisc}
\usepackage{tablefootnote}
\usepackage{adjustbox}
\usepackage[normalem]{ulem}
\usepackage{url}
\usepackage{multirow}
\usepackage{longtable}
\usepackage{lipsum} 
\usepackage{mathtools}
\usepackage[colorinlistoftodos]{todonotes}
\usepackage{microtype}
\usepackage[normalem]{ulem}

\def\BibTeX{{\rm B\kern-.05em{\sc i\kern-.025em b}\kern-.08em
		T\kern-.1667em\lower.7ex\hbox{E}\kern-.125emX}}

\let\oldIEEEkeywords\IEEEkeywords
\def\IEEEkeywords{\oldIEEEkeywords\itshape\ignorespaces}
\usepackage{caption}% http://ctan.org/pkg/caption
\newcommand{\ahm}[1]{{\color{blue}#1}}

\newcommand{\out}[1]{{\unskip}}
\newcommand{\ourapproach}{RCA}

\usepackage[bottom]{footmisc}
\SetAlgorithmName{Listing}{algorithmautorefname}{list of algorithms name}

\DeclarePairedDelimiter{\ceil}{\lceil}{\rceil}
\usepackage[bottom]{footmisc}
\usepackage{sistyle}
\SIthousandsep{,}
\definecolor{darkolivegreen}{rgb}{0.33, 0.42, 0.18}

\begin{document}
\title{RCA: A Resourceful Coordination Approach \\for Multilevel Scheduling} 
\author{Ahmed Eleliemy and Florina M. Ciorba
	University of Basel, Switzerland\\ firstname.lastname@unibas.ch
}
\date{}
\maketitle
\tableofcontents
\newpage
\setcounter{page}{1} 
\input{abstract.tex}
\input{introduction.tex}
%\input{background.tex}
\input{relatedwork.tex}

\input{multi_level_approach.tex}

\input{design_of_experiments.tex}

\input{experimental_evaluation.tex}

\input{conclusion_futurework.tex}
\newpage
\bibliographystyle{IEEEtran}
\bibliography{references}

\end{document}

%% file: abstract.tex
% !TEX root =  paper_index.tex
\begin{abstract}
HPC users aim to improve their execution times without particular regard for increasing system utilization.
On the contrary, HPC operators favor increasing the number of executed applications
per time unit and increasing system utilization. 
This difference in the preferences promotes the following operational model. 
Applications execute on exclusively-allocated computing resources for a specific time, and applications are assumed to utilize the allocated resources efficiently.	
In many cases, this operational model is inefficient, i.e., applications may not fully utilize their allocated resources. 
This inefficiency results in increasing application execution time and decreasing system utilization.
In this work, we propose a \emph{resourceful coordination approach} (\ourapproach{}) that enables the cooperation between, currently independent, batch- and \mbox{application-level} schedulers. 
\ourapproach{} enables application schedulers to share their allocated but idle computing resources with other applications through the batch system. 
%With enabling this coordination, \ourapproach{} avoids resource shrinking operations and associated performance penalties that are typical of dynamic resource and job management systems. 
%The current work employs a \mbox{Slurm-based} simulator (at the \mbox{batch-level}) and a \mbox{SimGrid-based} simulator (at the \mbox{application-level}) to implement the proposed approach. 
The effective system performance~(ESP) benchmark is used to assess the proposed approach. 
%Three \mbox{self-scheduling} techniques (static, guided \mbox{self-scheduling} and adaptive factoring) are used to execute jobs on their assigned resources. 
The results show that \ourapproach{} increased system utilization up to 12.6\% and decreased system makespan by the same percent without affecting applications' performance.

\end{abstract}
\newpage

%% file: introduction.tex
% !TEX root =  paper_index.tex
\section{Introduction}
\label{sec:intro}	
%\out{Scientists are eager to utilize additional computing resources to execute their applications to advance their understanding of various complex phenomena.
%This eagerness drives the rapid technological development in HPC.} 
Modern HPC systems exhibit parallelism at various hardware levels (node, socket, core, vector unit, etc.). %\out{\footnote{\url{https://www.top500.org/}}.} 
The efficient utilization of these levels of parallelism becomes more challenging with the increase of the degree of parallelism within each level.
When a \mbox{large-scale} HPC system wastes only 1\% to 10\% of its computing cycles, it wastes energy that could support a small city~\cite{sarood2014maximizing}.
In practice, HPC users aim to improve their execution times without particular regard for increasing the HPC system utilization.
This leads to the following operational model:  applications execute on a set of \mbox{exclusively-allocated} computing resources for a certain time (space policy), and applications are assumed to utilize the allocated resources efficiently via efficient domain decomposition and work assignment (balanced execution).

%Dynamic resource allocation (DRA) \ahm{plays} a significant role in increasing the usage efficiency of HPC systems~\cite{blazewicz2001approximation,kale2002malleable,Mounie99efficientapproximation,prabhakaran2015batch,utrera2004implementing}. 
Dynamic resource allocation (DRA) plays a significant role in increasing the usage efficiency of HPC systems~\cite{Mounie99efficientapproximation,prabhakaran2015batch}. %blazewicz2001approximation,kale2002malleable,utrera2004implementing
%DRA  means that the number of allocated resources of a given job varies during job execution.
%DRA can be driven by  system, jobs, or both~\cite{feitelson1996toward,feitelson1997theory}. 
A \emph{batch system}, also known as a resource and job management system~(RJMS), is the middleware responsible for resource allocation (RA) and for managing job execution on the allocated resources~\cite{georgiou2012evaluating}.
A batch system supports DRA if it implements scheduling policies that can increase or decrease the number of allocated resources of a given job.
%Expanding or shrinking running jobs is beneficial in many scenarios.
%For instance, when one job is not fully utilizing its allocated resources, a batch system can shrink this job and allows other jobs to exploit the freed resources.
%In this scenario, jobs need to smoothly adapt to new resource allocations (malleable jobs)~\cite{feitelson1996toward,feitelson1997theory}.
A \emph{job} is an instance of an \emph{application} that requires a certain amount of computational resources to execute on.
\emph{Evolving jobs} can request more or fewer resources than their current allocation based on the evolution of their computational load during the execution~\cite{feitelson1997theory,feitelson1996toward}.
\emph{Malleable jobs} need to adapt to new resource allocations smoothly~\cite{feitelson1997theory}.
%However, batch systems, in this scenario, need scheduling mechanisms and policies to respond to job expansion/shrink requests.
In practice, production RJMS systems, such as Slurm~\cite{yoo2003slurm} and Torque~\cite{staples2006torque}, support static resource allocation (SRA), where the number of the allocated resources is defined before the job execution and cannot subsequently be changed. 
%Certain research efforts~\cite{compres2016infrastructure,d2018drom,d2019holistic,prabhakaran2015batch,prabhakaran2014batch} extended production RJMS, such as Slurm~\cite{yoo2003slurm} and Torque~\cite{staples2006torque}, to support job malleability.
%Certain research efforts~\cite{compres2016infrastructure,d2018drom,d2019holistic,prabhakaran2015batch,prabhakaran2014batch} \ahm{have} extended production RJMS, such as Slurm~\cite{yoo2003slurm} and Torque~\cite{staples2006torque}, to support job malleability.
Certain research efforts have extended production RJMS to support malleability~\cite{compres2016infrastructure,d2019holistic}.%prabhakaran2014batch,prabhakaran2015batch,d2018drom
%Other research efforts studied DRA and via simulation~\cite{carroll2010incentive,sun2011fair,Mounie99efficientapproximation}.

%removed for space limits
%The essential challenge of supporting DRA \ahm{lies in} the programming paradigms of the running jobs.
%Modern programming paradigms, such as Charm++~\cite{kale1993charm} and OmpSs~\cite{bueno2011productive}, are adaptive and offer mechanisms to simplify the support of malleability.    
%However, \ahm{the} message passing interface MPI~\cite{MPIForum} is the dominant \ahm{multiprocess} programming approach for HPC applications, and the latest standard~(MPI~3.1) does not support malleability.
%Furthermore, the \ahm{existing extensions of certain MPI implementations require} scientists to reprogram their applications to support malleability~\cite{prabhakaran2014batch}.
%\ahm{Other} research efforts proposed specific approaches to bring malleability \ahm{into} MPI applications with minimal or without programming efforts~\cite{el2007dynamic,el2009malleable,mo2017large,utrera2004implementing}, but these approaches introduced certain overhead to the applications. 

Node sharing can also significantly improve system utilization via the simultaneous execution of multiple applications on the \emph{same} computing node~\cite{simakov2018slurm}.
The main challenge to achieve efficient node sharing is to identify applications that do not share the same set of resources.
Node sharing is much easier to implement than DRA. \out{due to the fact that} 
Node sharing does not require any changes or support from  applications and/or their underlying programming paradigms. 
Early research efforts introduced the PARbench which is a benchmark that assesses the performance impact of running mutiple jobs in a multiprogramming environment~\cite{Na91,Na2006}. 
The results showed  that the performance impact varies from being significant to be minor  based on the configuration of the system and the job requirements. 
This large variation makes node sharing not a common approach in practice, i.e., HPC users mainly have performance concerns that may arise from sharing \mbox{node-level} resources.  

The current work introduces a \emph{resourceful coordination approach} (\ourapproach{}) to increase system utilization via coordination between batch and application schedulers.
%The proposed approach is a part of a multilevel scheduling approach (MLS)~\cite{MLS}, where exchanging information across multiple levels of scheduling, achieves improved applications' performance and efficient system utilization.
\ourapproach{}  enables cooperation between the currently independent  batch and application schedulers. 
It enables application schedulers to share their allocated but idle computing resources with other applications through the batch system.
\textit{the proposed resourceful coordination approach is not an explicit dynamic resource allocation nor a node sharing approach but a unique approach that leverages the advantages of both of these approaches}. 
It \emph{offers an efficient idle resource sharing without shrink or expansion operations} on the application side.

To implement \ourapproach{} a \mbox{Slurm-based} simulator~\cite{simakov2018slurm} is employed as a batch scheduling framework together with a \mbox{SimGrid-based} simulator~\cite{mohammed2019approach} as an application scheduling framework.
%The source code of the \mbox{Slurm-based} simulator is actually the code of the Slurm RJMS with certain modifications.
Both \mbox{SimGrid-based} and \mbox{Slurm-based} simulators have been shown to be \emph{realistic} in terms of the close agreement between results obtained natively via direct experiments on HPC systems and results obtained via simulation~\cite{simakov2018slurm,mohammed2019approach}.

%The effective system performance~(ESP) is a \mbox{well-know} benchmark devised to evaluate \mbox{system-level} performance, including job scheduling efficiency~\cite{ESP1}.
To assess the usefulness of \ourapproach{}, we employ the effective system performance benchmark~(ESP)~\cite{ESP2}, and instantiate it with \emph{two workloads}.
The computational load of the jobs of the first workload is represented by a \mbox{computationally-intensive} parallel application called parallel \mbox{spin-image} algorithm (PSIA)~\cite{PSIA} form computer vision, while 
the second workload is represented by a \mbox{well-known} parallel kernel, the Mandelbrot set~\cite{Mandelbrot}.
For each job \out{in each of the two workloads}, three \mbox{application-level} scheduling techniques, called static~\cite{li1993locality}, guided \mbox{self-scheduling}~\cite{GSS}, and adaptive factoring~\cite{AF}), are used to balance applications' execution on their assigned resources. 
%This setup created a large set of experiments.
The experimental results showed that \ourapproach{} increased system utilization up to 12.6\% and decreased the system makespan by the same percent without affecting applications' performance.

This work makes the following contributions:
(1)~Introduces \ourapproach{} as a cooperation approach between batch and \mbox{application-level} schedulers.
(2)~Converts a \mbox{Slurm-based} simulator~\cite{simakov2018slurm} into an \mbox{event-based} simulator to evaluate the proposed approach. This extension yields deterministic and reproducible results.
(3)~Enables simulations of HPC workloads at fine (tasks within applications) and coarse (jobs within a workload) scales. 
To gain additional \mbox{in-depth} insights into the system and applications performance, we visualize the simulation events collected at both batch- and \mbox{application-level} by converting them to an \mbox{OTF2-based} trace that is compatible with trace visualization tools, such as Vampir~\cite{Vampir}.

The significance of the present work is that \ourapproach{} allows static RA to overcome the low system utilization, while avoiding the overhead of traditional DRA. 
The implementation extensions introduced in the Slurm simulator, convert it into an event-based simulator. This conversion is critical for simulations of high performance computing systems, as it delivers deterministic and results.

%Moreover, we collect simulation events (at both batch- and \mbox{application-level}) and convert them to an \mbox{OTF2-based} trace to use \emph{existing} trace visualization tools, such as Vampir~\cite{Vampir}, to visualize the simulation events and 
%(4)~Designs and implements a trace converter that enables the extended Slurm simulator to produce OTF2~\cite{OTF2} traces. 
%Hence, standard trace visualization tools, such as Vampir~\cite{Vampir}, can be used to visualize the output traces and to enable in depth visual analysis.

The remainder of the paper is organized as follows.
Section~\ref{sec:relatedwork} provides the background on batch- and \mbox{application-level} scheduling.
The most relevant research efforts are surveyed and reviewed in Section~\ref{sec:relatedwork}.
The  proposed resourceful coordination approach~(\ourapproach{}) is presented in Section~\ref{sec:proposed}, with details about the extensions introduced to existing batch and application simulators. 
In Section~\ref{sec:doe}, the evaluation methodology and results are presented and discussed.
The paper concludes and outlines directions for future work in Section~\ref{sec:conclusion}.  

%% file: relatedwork.tex
% !TEX root =  paper_index.tex
\section{Background and Related Work}
\label{sec:relatedwork}
\textbf{Background.} 
RJMS employ various batch scheduling techniques, such as the \emph{\mbox{first-come-first-serve}}~(FCFS) which schedules next the job with the earliest arrival time.
In practice, HPC system administrators employ a simple configuration of FCFS with backfilling~(BF)~\cite{BF}.
Backfilling is a supporting scheduling technique that helps to increase system utilization by executing small jobs (which request a small number of computing resources) when there are insufficient available computing resources to assign to the highest priority jobs in the queue. 

At the application level, we consider three application level scheduling techniques: static~(STATIC)~\cite{li1993locality}, guided self-scheduling~(GSS)~\cite{GSS}, and adaptive factoring~(AF)~\cite{AF}.
STATIC~\cite{li1993locality}, also known as straightforward parallelization, assigns each computing resource a chunk of loop iterations (or tasks) equal to $\ceil{N/P}$, where $N$ and $P$ are the total number of loop iterations and the total number of computing resources, respectively.
GSS is a dynamic \mbox{self-scheduling} technique that uses a \mbox{non-linear} function to self-schedule a decreasing chunk sizes.
At every scheduling step, GSS divides the remaining loop iterations by the total number of processing elements.
AF~\cite{AF} is an adaptive \mbox{self-scheduling} technique that is based on the factoring~(FAC) technique~\cite{FAC}.
FAC requires prior knowledge of the mean $\mu$ and the standard deviation $\sigma$ of the loop iterations execution times.
Unlike FAC, AF learns both $\mu$ and $\sigma$ for each computing resource during applications' execution to ensure full adaptivity to all factors that cause load imbalance.

\textbf{Related Work.} A notable research effort implemented an elastic execution framework for MPI applications~\cite{compres2016infrastructure}.
The framework introduced certain extensions to the MPI standard and to the Slurm RJMS. 
These extensions permit dynamic change of the number of processes of a given application in a way that addresses several challenges of the original dynamic process support of the MPI standard.
The extensions included four new MPI functions. %: (1)~\texttt{MPI\_INIT\_ADAPT} that replaces the \texttt{MPI\_INIT} and initializes the library in adaptive mode,
%(2)~\texttt{MPI\_PROBE\_ADAPT} that check whether the application needs to adapt, (3)~\texttt{MPI\_COMM\_APAPT\_BEGIN}, and (4)~\texttt{MPI\_COMM\_APAPT\_COMMIT} to start and completes adaptation window.
%The authors used an application that simulates a Tsunami to assess the performance of the elastic framework.
%In a tsunami simulation, the number of grid cells increases when wave propagate towards the coast, i.e., the application has a growing increase in the resource it needs over execution time.
%The elastic framework had a significant better resource utilization than the static MPI.
The elastic framework requires application scientists to use the new MPI functions to support application malleability.
%Such requirement could be a drawback or a limitation of the elastic MPI framework.
%A \mbox{large-scale} study that examined more than one hundred MPI applications showed that most of the MPI applications only use MPI 1.0 features~\cite{laguna2019large}.
%For instance, \mbox{non-blocking} collectives and neighborhood collectives are MPI 3.0 features and found to be in less than $1\%$ of the examined applications.
%The cost of rewriting working codes can be one of the reasons behind that fact.

This elastic MPI framework has the same goal as \ourapproach{}. 
However, \ourapproach{} shifts the responsibility of releasing or requesting computing resource to the application scheduler rather than the application code itself. 
Moreover, in \ourapproach{}, allowing one application to share idle computing resources with others does not require shrinking operation at the side of that application. 
This keeps the overhead low.

The dynamic resource ownership management~(DROM) is a recent research effort that allows RMJS to address the efficient resource usage challenge~\cite{d2018drom}.  
Compared to the elastic MPI framework~\cite{compres2016infrastructure}, the DROM APIs provide  effortless malleability for RMJS that requires no change in applications' source codes.
The DROM APIs exploit the finest level of parallelism to support application malleability, i.e., changing the number of the threads assigned to a computing resource to create a new room for other applications on the same computing resource.
One may use the DROM APIs with load balancing libraries similar to LeWI~\cite{LeWI} (LeWI is a runtime library that uses standard mechanisms, such as OMPT~\cite{OMPT} to monitor application execution.).
LeWI can enhance application performance and increase resource utilization of indvidual computing nodes.

%A holistic dynamic scheduling policy, called slowdown driven \mbox{(SD-policy)}~\cite{d2019holistic} was proposed based on the DROM APIs.
%The \mbox{SD-policy} applies \textit{backfilling} by selecting small jobs to share nodes with other running jobs .
%The \mbox{SD-policy} depends on the DROM APIs to achieve efficient \mbox{node-sharing}.

The DROM APIs and the LeWI library are similar to \ourapproach{} in the current work in the sense that we address the challenge of efficient resource usage, while our target is to enable  cooperation between the scheduling of different applications via batch systems.
For instance,  waiting or running applications (need more computing resources) may communicate their needs to the RJMS, which requests other \mbox{MPI-based} applications to stop scheduling any workload on the required computing resources for a certain period of time.
In this scenario, the schedulers of different applications cooperate with each others through the RJMS.
When an application scheduler decides not to schedule any workload on a certain process,  the process can be entirely suspended by the operating system and their computing resource can be used by other applications.

Other research efforts that are relevant to the present work include the recent advancements in the Slurm Simulator~\cite{lucero2011simulation}.
There are two distinctive versions of the Slurm Simulator Slurm V1~\cite{jokanovic2018evaluating,rodrigo2017scsf,trofinoff2015using} and Slurm V2~\cite{simakov2017slurm,simakov2018slurm}.
Slurm V2 is extensively simplified compared to Slurm V1, i.e, Slurm V2 \textit{serializes} the code on a single process, called \textit{sim\_controller}.
Such a simplification can been seen as a disadvantage because the simulator loses certain features, e.g., plugins that are used inside a Slurm node.
However, the same simplification can also be seen as an advantage when the target simulation scenarios do not use nor depend on these feature. 
This yields a clear code that is easier to extend, maintain, and debug.
Slurm V2 comes with a detailed documentation on how to reuse and extend. %\footnote{\url{https://github.com/ubccr-slurm-simulator/slurm_sim_tools/blob/master/\\doc/slurm_sim_manual.pdf}}.
Therefore, the present work uses and extends Slurm V2. 
\ourapproach{} is independent of  the simulator and the same extensions can also be implemented in Slurm V1.

%% file: multi_level_approach.tex
% !TEX root =  paper_index.tex
\section{Resourceful Coordination Approach for Multilevel Scheduling}
\label{sec:proposed}	
The resourceful coordination approach~(\ourapproach{}) requires information exchange between batch and application level schedulers: 
 (1)~From the application schedulers to the batch scheduler.
The application schedulers report the status of their free computing resources and the remaining amount of work.
(2)~From the batch scheduler to the application schedulers. 
The batch scheduler can take advantage of knowing the execution history of certain applications and can benefit from additional hints that the user may provide, such as expected applications' execution time, \mbox{communication/computation} ratio, etc.
The information exchange allows the batch scheduler to reuse computing resources as soon as they become idle, and there are no more tasks from the job that can be assigned to them.
User hints allow the batch scheduler to identify applications that experience minimal performance degradation when they exclude a specific number of their allocated resources.
The exclusion means that the application schedulers will not schedule further tasks on the excluded resource. 
This exclusion differs from shrinking the resource allocation of malleable jobs. 
In \ourapproach{}, the application still owns the temporarily relinquished computing resource, but it allows other applications use it. 
\ourapproach{} allows application schedulers to accept or reject resource exclusion requests from the batch scheduler. 

%In the first case, the application schedulers know information regarding current scheduled and remain workload. Therefore, they can decide which resource to offer for other application through the batch system.
%In the second case, the batch system knows historical information regarding all application, and it can select which application can offer computing resource to others without or with minimum performance loss.

% \begin{figure*}[!t]
 %	\centering
 %	\includegraphics[clip,trim=0cm 0cm 0cm 0cm,scale=0.35]{figures/removing_resource}
 %	\caption{Illustration that shows  }
% 	\label{fig:rm}
% \end{figure*}
 \begin{figure}[!b]
 	\centering
 	\renewcommand{\figurename}{Figure}
 	\includegraphics[clip,trim=0cm 0cm 0cm 0cm, width=0.7\columnwidth]{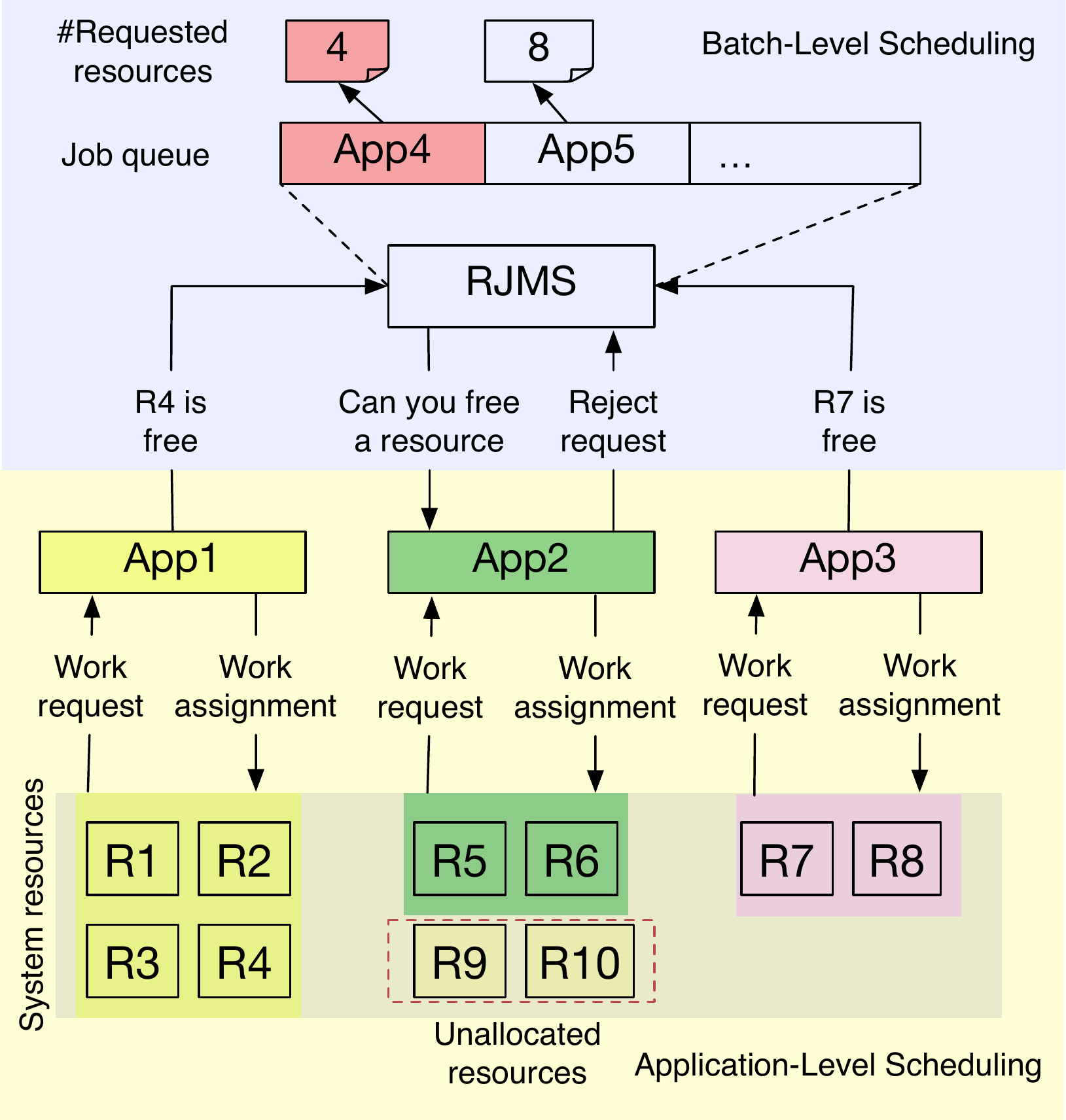}
 	\caption{Proposed resourceful coordination approach (\ourapproach{}) in which applications (e.g., App1, App4) cooperate by yielding idle resources (e.g., R4) through the batch system.}
 	\label{fig:mls}
 \end{figure}

%The idea of the proposed approach facilitates the cooperation between all levels of scheduling via information exchange.
Figure~\ref{fig:mls} illustrates three executing applications (App1, App2, App3) and two queued applications (App4 and App5).
First-come-first-serve~(FCFS) is employed at the \mbox{batch-level} to schedule the five jobs.
App4 has higher priority than App5.
App4 requests four computing resources and only two resources are available: R9 and R10.
However, the batch system cannot start App4 due to  insufficient free resources.
\out{Assuming that BF is enabled, then the batch system launches any job from the queue that requests at most two resources.
In the example, App5 requests eight resources, and no other applications exist in the queue.}
In this case, existing batch scheduling systems would leave App4 waiting in the queue and R9 and R10 idle until one of the executing applications finish.
In contrast, in \ourapproach{}, the batch system receives information form application schedulers during applications' execution.
App1 and App3 report that R4 and R7 became free.
R4 and R7 can be reassigned to other applications through the  batch system.
The information from App1 and App3 may be reported at different times.   
Once the batch system receives these two reports, and if App4 is still in the queue, the batch system can assign R4, R9, R10, and R7 to App4 which can then begin execution.

The batch system can identify (based on applications' execution history) applications that can relinquish certain resources without performance degradation.
In the example illustrated in Figure~\ref{fig:mls}, the batch system identified App2 as such an application. 
The application scheduler of App2 rejected the request and did not release any resources.
In \ourapproach{}, the batch scheduler does not control the ALS decisions. 
Application schedulers can reject the release of resource requests. 
Accepting or rejecting batch requests can be seen as a higher level of cooperation than reporting resource idle time that can be enabled or disabled based on users' preferences.
Moreover, the batch system leaves the decision regarding which resource to be freed to the application scheduler.

\ourapproach{} separates concerns between BLS and ALS.
BLS provides the required number of resources to waiting jobs, while ALS decides which resource(s) is (are) ready to be released right away.
This separation eliminates BLS' need to employ techniques that extrapolate and predict future resource requirements for executing applications. 
ALS schedulers always have information about the remaining computational workload.
With RCA, this information can easily be shared with BLS. 

%Unfortunately, the existing scheduling libraries and batch systems do not allow us to develop and evaluate the proposed scheduling approach.
%As a first step, the current work implements and evaluates the usefulness of the aforementioned proposed approach in simulation.
%The current work exploits two simulation frameworks for batch and application level scheduling.
\textbf{Design details:}
%As a first step, the current work implements and evaluates the usefulness of the aforementioned proposed approach in simulation
At \mbox{batch-level} scheduling, the current work employs the widely used Slurm simulator~\cite{lucero2011simulation}.
We extend one of the latest versions of the Slurm simulator~\cite{simakov2018slurm} to support \ourapproach{}, modifications being listed in Listing~\ref{algo:slurm_sim}. 
%These advancements lead to two distinctive versions V1~\cite{jokanovic2018evaluating,rodrigo2017scsf,trofinoff2015using} and V2~\cite{simakov2017slurm,simakov2018slurm}.
%The main difference between is that V2 is  extensively simplified compared to  V1, i.e, V2 \textit{serializes} the code on a single process, called \textit{sim\_controller}.
%Such a simplification can been seen as a disadvantage because the simulator losses certain features, e.g., plugins that are used inside SLURM node.
%However, the same simplification can also be seen as an advantage when the target simulation scenarios do not use or depend on these feature. 
%The simplification yields a clear code that is easier to extend, maintain, and debug.
%Moreover, V2 has a detailed documentation on how to reuse and extend\footnote{\url{https://github.com/ubccr-slurm-simulator/slurm\_sim\_tools/blob/master/\\doc/slurm\_sim\_manual.pdf}}.

 \begin{algorithm}[!t]
\textbf{slurm\_sim\_controller()}\{\\
\setcounter{AlgoLine}{0}
 \ShowLn	
 read\_slurm\_sim\_configuration(sim\_config)\;

 {\color{black}  \ShowLn {extract\_als\_configuration(als\_config)}}\; %darkolivegreen
 
 {\color{black} \ShowLn sim\_read\_job\_trace(trace\_head)}\; %blue
 
 {\color{black} \ShowLn synchronize\_with\_app\_simulator(als\_config)}\; %darkolivegreen
 \ShowLn
  \While{True}
  {
  	
  	{\color{black} \ShowLn run\_scheduling\_round();} %darkolivegreen \texttt{/*Listing~2*/}\\
  	
  	{\color{black} \ShowLn update\_SimGrid\_simulation\_clock()\; %darkolivegreen
  		\ShowLn
  	\If{no\_jobs\_to\_submit()}
  	{
  		\ShowLn
  		\If{no\_running\_apps()}
  		{
  			\ShowLn
  			collect\_simulation\_trace()\;
  			\ShowLn
  			end\_app\_simulator()\;
  			\ShowLn
  			exit()\;
  		}
  	}}
  	{\color{black} \ShowLn sim\_submit\_jobs()\; %blue
  	\ShowLn
  	sim\_process\_finished\_jobs()\;
  	\ShowLn
  	sim\_cancel\_jobs()\;
  	\ShowLn
  	sim\_schedule()}\;
	 \ShowLn
	 sim\_run\_priority\_decay()\;
  	\ShowLn
  	schdeule\_plugin\_run\_once()\;
  	\ShowLn
  	sim\_sinfo()\;
  	\ShowLn
  	sim\_squeue()\;
  }
  \} %\texttt{/*{\color{black} original}, {\color{black} new}, {\color{black} modified} code*/}
 \caption{\mbox{Batch-level} scheduling}
 \label{algo:slurm_sim}
 \end{algorithm} 

Listing~\ref{algo:slurm_sim}, Line~2 shows the new code we added to allow the Slurm simulator to read ALS information, such as the ALS scheduling method.
Line~3 represents the modified code that extends the Slurm simulator to accept workloads in the standard workload format (SWF)~\cite{PWA}.  
This modification enables the simulation of various workloads from production HPC systems that are available in the public workload archive~\cite{PWA}.
Lines~4 to~12 represent new added code that connects the \mbox{SimGrid-based} simulator with the Slurm simulator.
Hence, the \mbox{SimGrid-based} simulator  works as an \textit{internal clock} for the Slurm simulator.
SimGrid simulations are \mbox{event-based} simulations, and consequently, the simulation time is only advanced by the occurrence of simulation events. 
In our approach, the simulation time is advanced only when scheduling events happen at either the batch- or \mbox{application-level}.
Lines~13 to~16 represent certain functions of the original Slurm simulator~\cite{simakov2018slurm} that we extended to produce or consume scheduling events of the \mbox{SimGrid-based} simulator.  

The communication between the two simulators employs a shared data structure called \emph{all\_apps}, which holds all information about jobs' execution (Line~1 in Listing~\ref{algo:sim}).
Scheduling events, such as starting a job on a specific set of resources, are produced by the \mbox{Slurm-based} simulator and stored in the \emph{all\_apps} data structure.
Also, scheduling events, such as job completion, are produced by the \mbox{SimGrid-based} simulator and are stored in the \emph{all\_apps} data structure.
Each simulator \textit{consumes} the events \emph{produced} by the other simulator.

%Collecting simulation events (at batch and application levels) and converting them to an \mbox{OTF2-based} trace allows us to employ \emph{existing} trace visualization tools, such as Vampir~\cite{Vampir}, to visualize the simulation events and to gain in-depth insights.

%\subsection{ALS simulation}
At \mbox{application-level} scheduling, the present work designs and extends an accurate \mbox{SimGrid-based} simulator~\cite{mohammed2019approach} that is used to simulate applications' executions with various DLS techniques by \textit{simultaneously} simulating the execution of several applications running on the same simulated HPC platform.
The intention behind this difference is to let the simulator account for application interference.
Earlier research efforts~\cite{eleliemytwolevel,mohammed2019approach} focused on the study of applications' performance under various scheduling techniques. 
In contrast, the current work relaxes the assumption of applications executing on separate sets of resources during their entire execution, thereby increasing the realism of the simulation.

%Application schedulers offer a subset of their resources back to the batch system to execute other applications.
%Unlike earlier research effort where applications were simulated simultaneously by multiple parallel instants of the simulator, the proposed simulator lose this feature.
Listing~\ref{algo:sim} shows a single \textit{scheduling round} of our extended \mbox{SimGrid-based} simulator.
A scheduling round refers to a scanning procedure where all simulated applications and their assigned resources are examined to identify the idle resources and to \mbox{self-schedule} the remaining work.
Listing~\ref{algo:sim} illustrates the logic of the function \textit{run\_scheduling\_round()} of Listing~\ref{algo:slurm_sim}.

%For instance, functions of Algorithm~\ref{algo:slurm_sim}, such as sim\_submit\_jobs(), sim\_process\_finished\_jobs(), and sim\_schedule(), access this dynamic list to add and schedule new jobs and to identify finished jobs or idle computing resources.}
\begin{algorithm}[!t]
	\textbf{run\_scheduling\_round()}\{\\
	\setcounter{AlgoLine}{0}
	\ShowLn	
	\ForEach{app in all\_apps}
	{
		\ShowLn
  unscheduled = check\_unscheduled\_tasks(app)\;
  \ShowLn
  hosts = get\_free\_hosts(app)\;
  \ShowLn
		\ForEach{host in hosts}
		{
			\ShowLn
			\If{unscheduled \textgreater 0}
			{
				\ShowLn
				scheduling\_method= schedudling\_method(app)\;
				\ShowLn
				tasks=chunk\_size(app, scheduling\_method)\;
				\ShowLn
				schedule\_tasks(host, tasks)\;
				\ShowLn
				unscheduled = unscheduled - tasks\;
			}
			\ShowLn
			release\_host(host,app)\;
		}
	}
	\}	\small{\texttt{/*scheduling round in SimGrid*/}}
	\caption{Application-level scheduling}
	\label{algo:sim}
\end{algorithm}

For native Slurm RJMS, the \mbox{BLS-ALS} communication can be implemented via remote procedure call (RPC) similar to the communication between the Slurm daemons (slurmctl and slurmd). 
The Slurm daemons periodically exchange messages to monitor resources' status.
These small messages have minimal impact on the performance of the running application.
The \mbox{BLS-ALS} communication are not periodic and they are occasionally sent.
For instance, \mbox{BLS-ALS} communication messages are sent when the originating entity is not executing any workload. 
The \mbox{BLS-ALS} communication messages in that sense will not degrade applications' performance.

%\subsection{ALS and BLS integration}
%\lipsum[2-3]

%The aforementioned scenario poses certain questions, such as \textit{what happens when App\_1 needs to reuse the release resource R6?} 
%\textit{Why does the batch scheduler select App\_2 to request resources?} 
%\textit{What happens if App\_4 needs more resource than released by App\_2?} 

%% file: design_of_experiments.tex
% !TEX root =  paper_index.tex
\section{Experimental Design and Evaluation}
\label{sec:doe}	
%\subsection{Design of experiments}
\textbf{Experimental design:} In all experiments reported herein, a simulated platform with 256 compute hosts is used. 
%Each of the hosts has a processor that contains 16~compute cores. 
A \mbox{fully-connected} network topology is used to connect all hosts.
The network fabric is assumed InfiniBand like with link bandwidth and latency of 50~Gbps and 500~ns, respectively.

The effective system performance (ESP)~\cite{ESP2} benchmark is used to evaluate the usefulness of the proposed approach.
ESP gives a description of batch workloads that can be used to assess batch systems' performance.
The description includes guidelines regarding the total number of jobs, estimated job execution time, number of requested resources per job, and job arrival times~\cite{ESP2,prabhakaran2015batch,georgiou2012evaluating}. 
Table~\ref{tab:esp} illustrates the characteristics of the ESP system benchmark, which consists of 230 jobs divided into 14 job categories.
Jobs of different categories require various numbers of computing resources, from 3.12\% to 100\% of the available computing resources. 
For instance, one job in Category A requires 8 computing resources (3.12\% of the entire system), while one job in Category Z requires  256 computing resources (the entire system).
%The arrival of the two jobs shapes the 
%, the full configuration jobs must be run before any further
%jobs are launched. The first full configuration job can only be submitted after
%10% of the theoretical minimum time has elapsed such that it is non-trivial to
%schedule. Similarly, the second full configuration job must complete within 90%
%of the test and is not simply the last job to be launched
%The first full configuration job can only be submitted after 10% of the theoretical minimum time has elapsed such that
% it is non-trivial to schedule. Similarly, the second full configuration job must complete within 90% of
% the test and is not simply the last job to be launched. The requirement to run these two full configuration
% jobs is a difficult test for a scheduler, but it is nonetheless a common scenario in capability environments.	 
 \begin{table}[!t]
 	\caption{Characteristics of the two implemented versions of the ESP system benchmark: \mbox{ESP-PSIA} and \mbox{ESP-Mandelbrot}.}
 	\centering
 	\begin{tabular}{r|r|r|r|r}
 		\multicolumn{1}{r|}{\begin{tabular}[c]{@{}r@{}}Category\\ ID\end{tabular}} & \multicolumn{1}{r|}{\begin{tabular}[c]{@{}r@{}}Requested\\  Hosts\end{tabular}} & \multicolumn{1}{r|}{\begin{tabular}[c]{@{}r@{}}Total\\  Jobs\end{tabular}} & \multicolumn{1}{r|}{\begin{tabular}[c]{@{}r@{}}ESP-PSIA\\ \#images \end{tabular}} & \multicolumn{1}{r}{\begin{tabular}[c]{@{}r@{}}ESP-Mandelbrot\\ \#iterations \end{tabular}} \\ \hline
 		A                                                                           & 8                                                                               & 75                                                                         & 32 K                                                                        & 0.635 M                                                                           \\
 		B                                                                           & 16                                                                              & 9                                                                          & 76.5 K                                                                      & 1.2 M                                                                             \\
 		C                                                                           & 128                                                                             & 3                                                                          & 800 K                                                                       & 15 M                                                                              \\
 		D                                                                           & 64                                                                              & 3                                                                          & 582 K                                                                       & 8.5 M                                                                             \\
 		E                                                                           & 128                                                                             & 3                                                                          & 595 K                                                                       & 8.8 M                                                                             \\
 		F                                                                           & 16                                                                              & 9                                                                          & 440 K                                                                       & 6.5 M                                                                             \\
 		G                                                                           & 32                                                                              & 6                                                                          & 635 K                                                                       & 1 M                                                                               \\
 		H                                                                           & 40                                                                              & 6                                                                          & 630 K                                                                       & 10 M                                                                              \\
 		I                                                                           & 8                                                                               & 24                                                                         & 170 K                                                                       & 3.35 M                                                                            \\
 		J                                                                           & 16                                                                              & 24                                                                         & 174.5 K                                                                     & 2.75 M                                                                            \\
 		K                                                                           & 24                                                                              & 15                                                                         & 172.6 K                                                                     & 2.85 M                                                                            \\
 		L                                                                           & 32                                                                              & 36                                                                         & 172.5 K                                                                     & 2.725 M                                                                           \\
 		M                                                                           & 64                                                                              & 15                                                                         & 176.5 K                                                                     & 2.65 M                                                                            \\
 		Z                                                                           & 256                                                                             & 2                                                                          & 375 K                                                                       & 5.25 M                                                                            \\ 
 	\end{tabular}
 	\label{tab:esp}
 \end{table}

\begin{figure}[!b]
	\centering
	\renewcommand{\figurename}{Figure}
		\includegraphics[width=0.8\columnwidth]{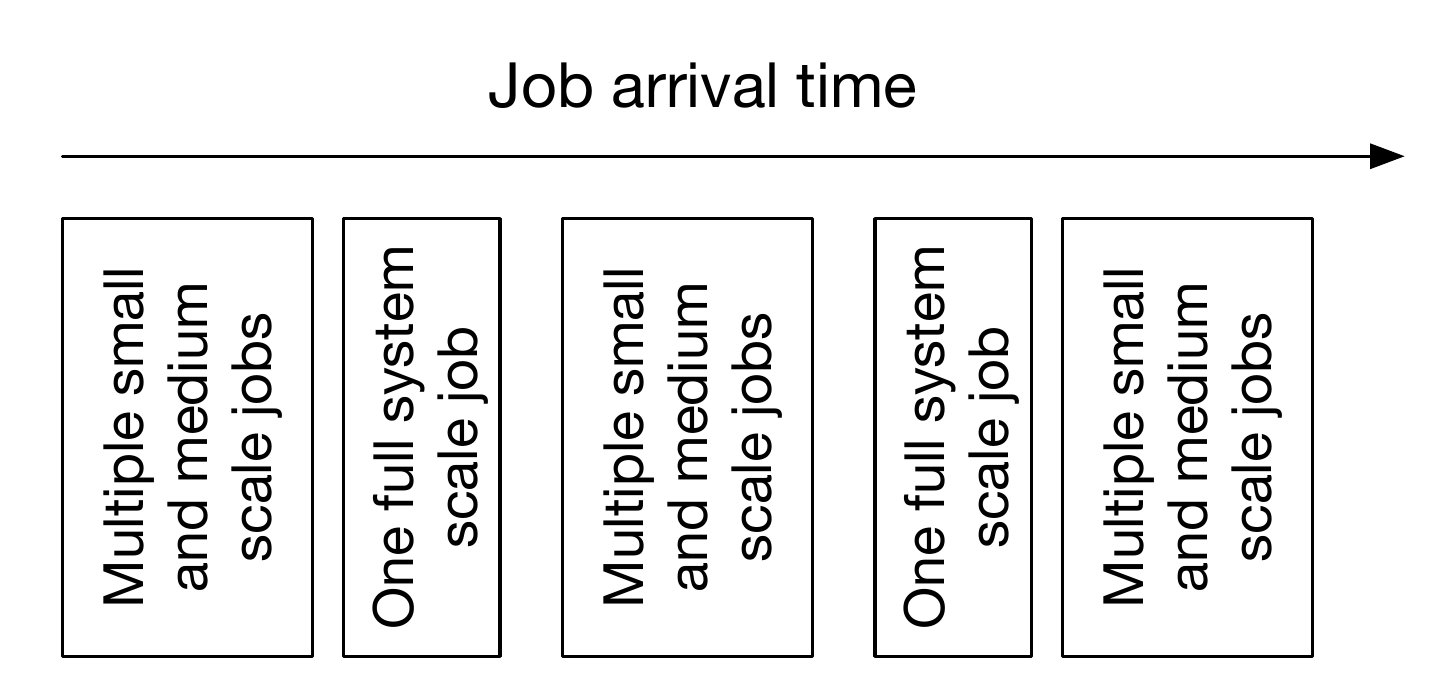}
	\caption{ESP job arrival scheme (adapted from~\cite{ESP1,ESP2}). Full system scale jobs refer to jobs from Category Z.  Multiple and small scale jobs refer to jobs from other categories.}
	\label{fig:espsub}
\end{figure}

Another essential factor in the ESP system benchmark is the job arrival time.
The ESP designers suggested a job arrival scheme in which Category Z jobs arrive in such a way that they divide the arrival timeline into 3 parts\out{(see Figure~\ref{fig:espd})}~\cite{ESP2}.  
This means that jobs arrive during the batch execution.
This arrival pattern prevents the batch scheduler from knowing the entire workload before the execution, which would be unrealistic.
Figure~\ref{fig:espsub} shows the job submission time for each job of the ESP. 
Once a \mbox{full-size} job (Job of Category Z) is submitted, no other jobs are submitted for a specific amount of time that is equal to 10\% of the ESP workload makespan.

\begin{figure}[!b]
	\centering
	\renewcommand{\figurename}{Figure}
	\begin{adjustbox}{minipage=0.7\linewidth}
		\centering
		\subfloat[ESP-PSIA]{%
			\includegraphics[width=\textwidth]{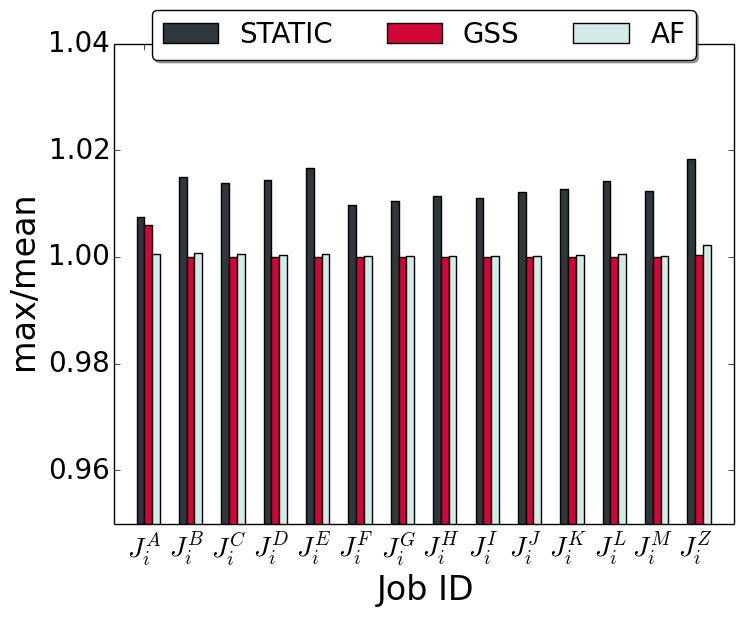}
			\label{fig:ESP-PSIA}%
		} 
		\\
		\subfloat[ESP-Mandelbrot] {%
			\includegraphics[width=\textwidth]{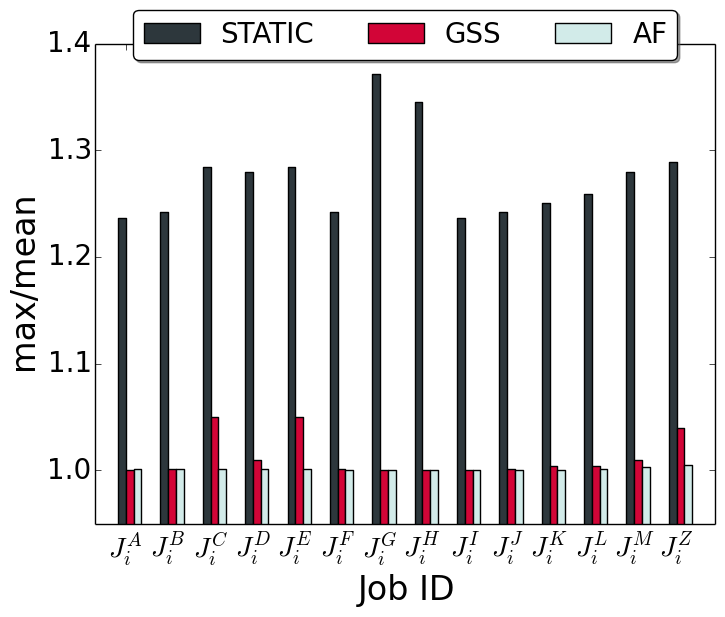}
			\label{fig:ESP-MLBT}%
		} 
	\end{adjustbox}
	\caption{Load imbalance profile of the jobs within the \mbox{ESP-PSIA} and \mbox{ESP-Mandelbrot} workloads. The ratio max/mean indicates the degree of balanced execution for each job $J^x_i$, where is $x$ is a job category (see Table~\ref{tab:esp}) and $i$ ranges according to the size of each job category. Values that are close to 1 denote a balanced execution.}
	\label{fig:ESP_PSIA-MLBT}
\end{figure}	

%\begin{figure}[!b]
%	\centering
%	\renewcommand{\figurename}{Figure}
%	\includegraphics[width=\columnwidth]{figures/espsub}
%	\caption{\ahm{Job submission time of the ESP workload. The submission gab represents almost 10\% of the ESP workload makespan.}}
%	\label{fig:espsub}
%\end{figure}
	
ESP jobs are synthetic and can be represented by various applications~\cite{ESP2}.
Here, we exemplify the ESP system benchmark with the PSIA and Mandelbrot applications.
We generate and use two different workloads of the ESP system benchmark called \emph{\mbox{ESP-PSIA}} and \emph{\mbox{ESP-Mandelbrot}}.
PSIA and Mandelbrot are chosen to represent two extremes of interest for testing our approach: a balanced execution (PSIA)  and a highly load imbalanced execution (Mandelbrot).
Moreover, individual research efforts~\cite{mohammed2019approach,Mohammed:2018a} proposed an accurate and verified representation of the computational workload of both applications in SimGrid.
A mixed workload that comprises both applications is planned as immediate future work.

PSIA~\cite{PSIA} is a \mbox{computationally-intensive} application from computer vision that consists of a large loop that dominates the entire execution.
Loop iterations in PSIA have different computational loads and require efficient loop scheduling to achieve a balanced execution of these iterations.
Various dynamic scheduling techniques can achieve a balanced execution for PSIA. Consequently, there are few differences in computing resource finishing times that execute the PSIA application.
Such times are important in this work as they represent idle resources that can be relinquished.
%This set of experiments forms the most challenging scenario for the proposed approach.
The Mandelbrot set is a \mbox{well-known} mathematical kernel.  
It contains a set of irregular and independent loops and has been used to evaluate DLS techniques in the literature~\cite{PLS,DSS}.

\begin{table}[!b]
	\centering
\begin{tabular}{|l|l|l|}
	\hline
	BLS workload & BLS technique & ALS technique \\ \hline
	ESP PSIA-based & \multirow{2}{*}{FCFS with BF} & \multirow{2}{*}{STATIC/ FAC/ AF} \\ \cline{1-1}
	ESP Mandelbrot-based &  &  \\ \hline
\end{tabular}

		\caption{ Details of the factorial experimental design for the performance evaluation of the proposed approach }
		\label{tab:doe}
\end{table}

The last two columns of Table~\ref{tab:esp} show the characteristics of the two versions of the ESP system benchmark workload that contain PSIA and Mandelbrot jobs.
Various input parameters control the execution of PSIA and Mandelbrot~\cite{PSIA,Mandelbrot}.
One parameter for each application is changed to let the applications meet the job execution category of the ESP~\cite{prabhakaran2015batch}.
For PSIA, \textit{\#images} indicates the total number of generated \mbox{spin-images}. 
For Mandelbrot, \textit{\#iterations} indicates the maximum number of iterations per pixel.
The two parameters are chosen because they had a linear relation to the application execution time. 
Therefore, it is more precise to estimate their initial values that meet the job execution category.

Figure~\ref{fig:ESP_PSIA-MLBT} shows the load imbalance profile of the two versions of the ESP system benchmark: \mbox{ESP-PSIA} and \mbox{ESP-Mandelbrot}.
The metric \emph{max/mean} denotes the ratio between the finishing time of the latest computing resource and the average finishing time of all computing resources that execute a certain job.
When the ratio \emph{max/mean} of a certain job is very close to one, the job has a balanced load execution on its allocated resources.

%% file: experimental_evaluation.tex
% !TEX root =  paper_index.tex
%\subsection{Results and discussion }
%\label{sec:discussion}	

\out{In Figure~\ref{fig:ESP-PSIA}, for all job categories, the values of \mbox{max/mean} are close to one.
This reflects the balanced load execution of the PSIA.
For \mbox{ESP-PSIA}, AF achieved the most balanced execution compared to GSS and STATIC (see Figure~\ref{fig:ESP-PSIA}).
AF also achieved a \mbox{fully-balanced} execution for \mbox{ESP-Mandelbrot} compared to STATIC and GSS (see Figure~\ref{fig:ESP-MLBT}).
The results in Figure~\ref{fig:ESP_PSIA-MLBT} indicate less idle resources when executing \mbox{ESP-PSIA} than when executing \mbox{ESP-Mandelbrot}. 
Therefore, the \mbox{ESP-PSIA} workload represents a challenging case for the proposed approach, i.e., computing resources have short idle times that can only briefly be exploited by other applications.}

\textbf{Experimental Evaluation and Discussion:} System utilization~(SU) is an important metric that indicates the efficiency of batch scheduling techniques.
We calculate SU as shown in Eq.~\ref{eq:su}, where  $T_k$ is the time that a computing resource $k$ spent executing jobs, $P$ is the total number the computing resources,  and $T_{batch}$ denotes the system makespan measured as the total execution time of \emph{the entire batch}, i.e., $T_{batch}= T_{j} - T_{i}$, where $T_{i}$ is the time when the first job starts execution and $T_{j}$ is the time when the last job in the batch completes execution.
System utilization ranges from 0\% to 100\%. 
Higher values of system utilization indicate better system performance.
\begin{equation}
\label{eq:su}
SU =  \frac{\sum_{k=0}^{P-1} T_k}{P * T_{batch}} * 100
\end{equation}

\begin{figure*}[!h]
	\centering
	\renewcommand{\figurename}{Figure}
		\centering
		\subfloat[FCFS + BF (without \ourapproach{})]{%
			\includegraphics[width=0.5\textwidth]{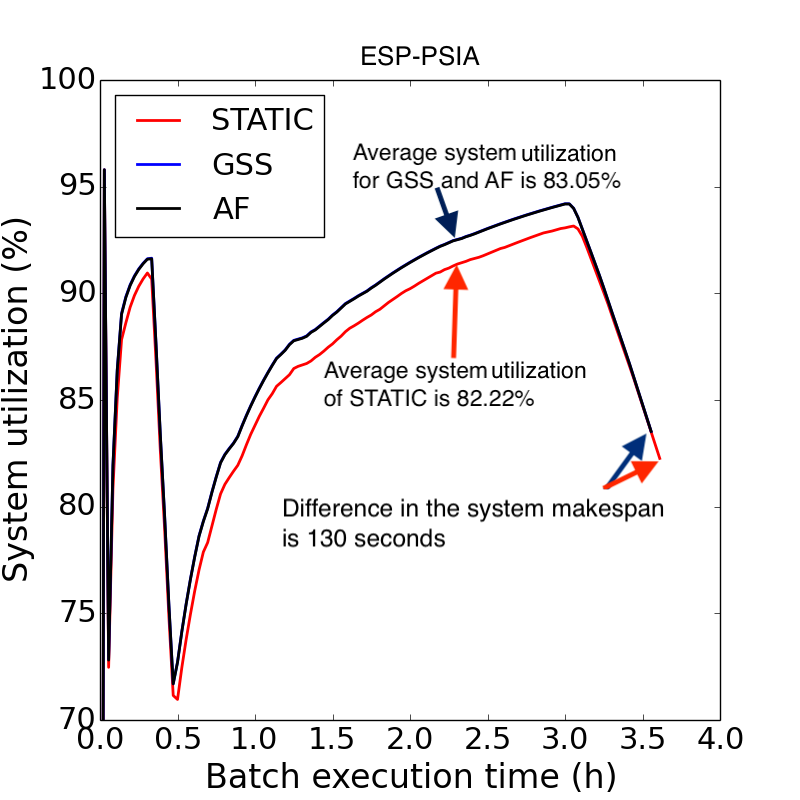}
			\label{fig:SU-PSIA-wout}%
		} 
		\subfloat[FCFS + BF + \ourapproach{}] {%
			\includegraphics[width=0.5\textwidth]{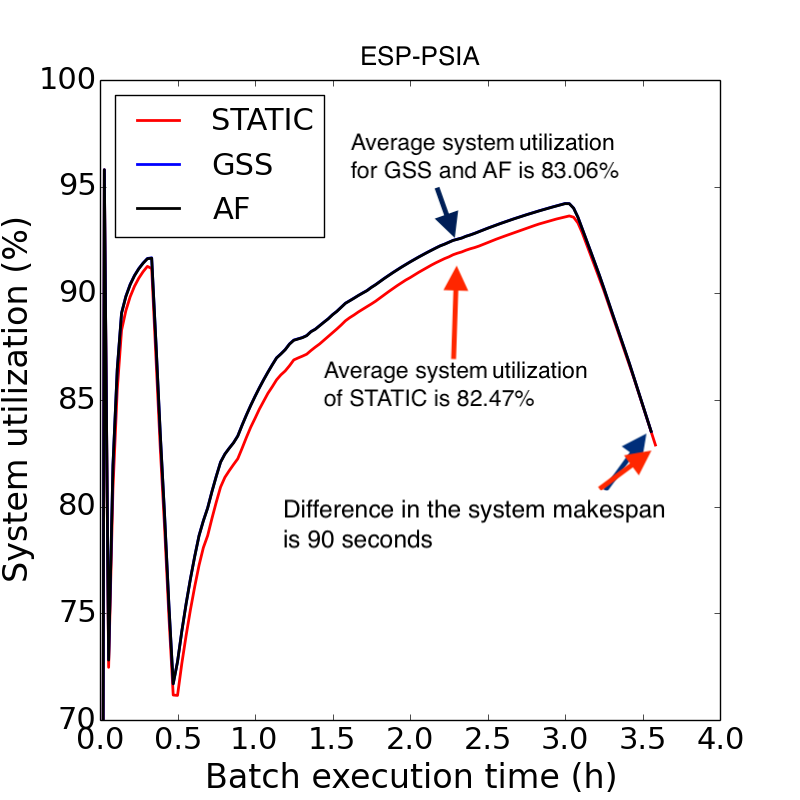}
			\label{fig:SU-PSIA-with}%
		} 
	\caption{System utilization of \mbox{ESP-PSIA} }
	\label{fig:SU-PSIA}
\end{figure*}

Figure~\ref{fig:SU-PSIA} shows the system utilization over batch execution time for the \mbox{ESP-PSIA} \textit{with} and \textit{without} the proposed approach. 
When our resourceful scheduling approach is not enabled in the simulation, the makespan of the \mbox{ESP-PSIA} using STATIC, GSS, and AF is \num{13000}, \num{12875}, and \num{12875} seconds, respectively (see Figure~\ref{fig:SU-PSIA-wout}). 
This corresponds to the increase in the system utilization in Figure~\ref{fig:SU-PSIA-wout}; the GSS and AF curves are slightly higher than the STATIC curve. 

Figure~\ref{fig:SU-PSIA-with} shows that the system makespan improved with our resourceful scheduling approach.
For instance, the system makespan for \mbox{ESP-PSIA} with STATIC is \num{12965} instead of \num{13000} seconds. 
For GSS and AF the improvement is not impressive.
As mentioned earlier, \mbox{ESP-PSIA} is an extreme case of a highly balanced execution.
This means that the differences in resource finishing times that execute the PSIA application are minimal.
In this case, \ourapproach{} has limited advantage.
%\mbox{ESP-PSIA} is challenging, and in practice, achieving a balanced load execution for all the jobs within a workload is infeasible in many cases.
One can still notice that the gap in system utilization when using STATIC, GSS, and FAC with \ourapproach{} (see 
Figure~\ref{fig:SU-PSIA-with}) is slightly smaller than the gap in Figure~\ref{fig:SU-PSIA-wout} that is without \ourapproach{}.
\out{Specifically, the gap has been decreased by $\frac{83.06-82.47}{83.05-82.22} = 71\%$.}

\begin{figure}[!h]
	\centering
	\renewcommand{\figurename}{Figure}
		\centering
		\subfloat[FCFS + BF (without \ourapproach{})]{%
			\includegraphics[width=0.5\textwidth]{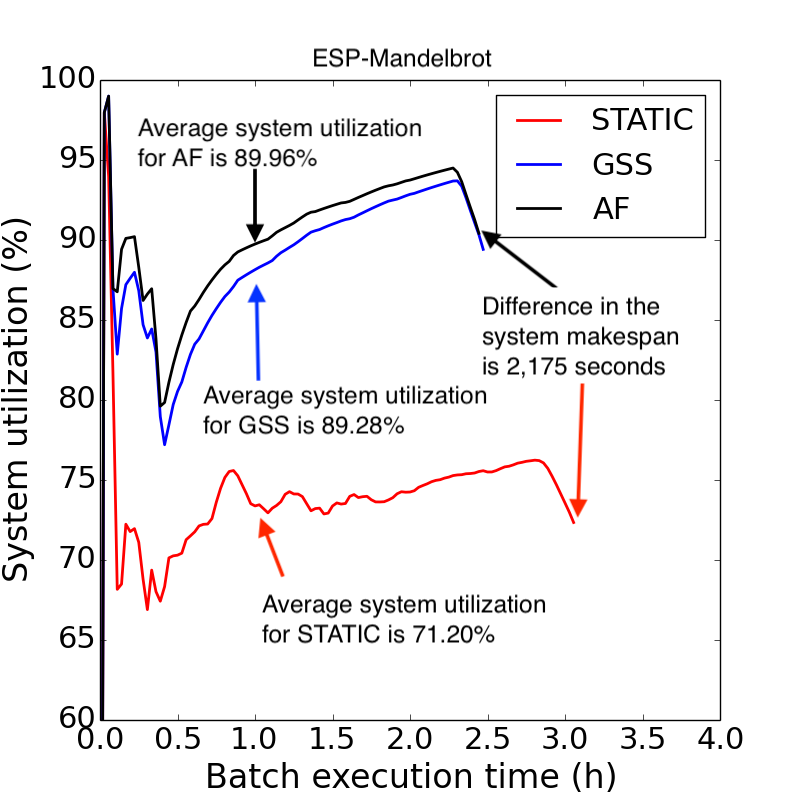}
			\label{fig:SU-MLBT-wout}%
		} 
		\subfloat[FCFS + BF + \ourapproach{}] {%
			\includegraphics[width=0.5\textwidth]{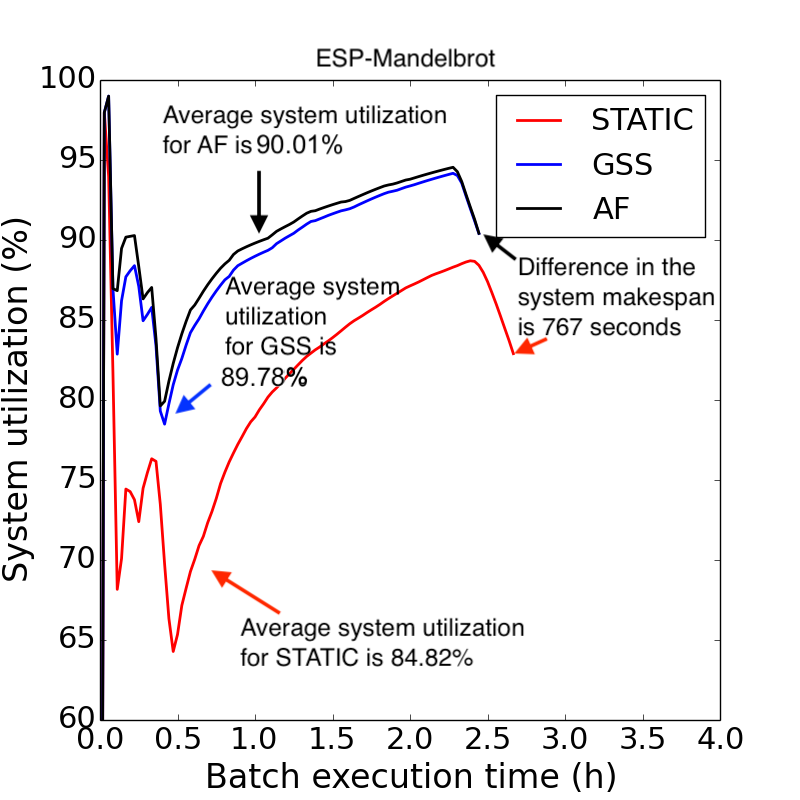}
			\label{fig:SU-MLBT-with}%
		} 
	\caption{System utilization of \mbox{ESP-Mandelbrot} }
	\label{fig:SU-MLBT}
\end{figure}

\begin{figure*}[!h]
	\renewcommand{\figurename}{Figure}
	\centering
	\subfloat[Poor system utilization without \ourapproach{} at 71.20\% due to idle resources,\newline while J7 approaches completion ]{%
		\includegraphics[width=0.8\textwidth, clip,trim=0cm 0cm 0cm 0cm]{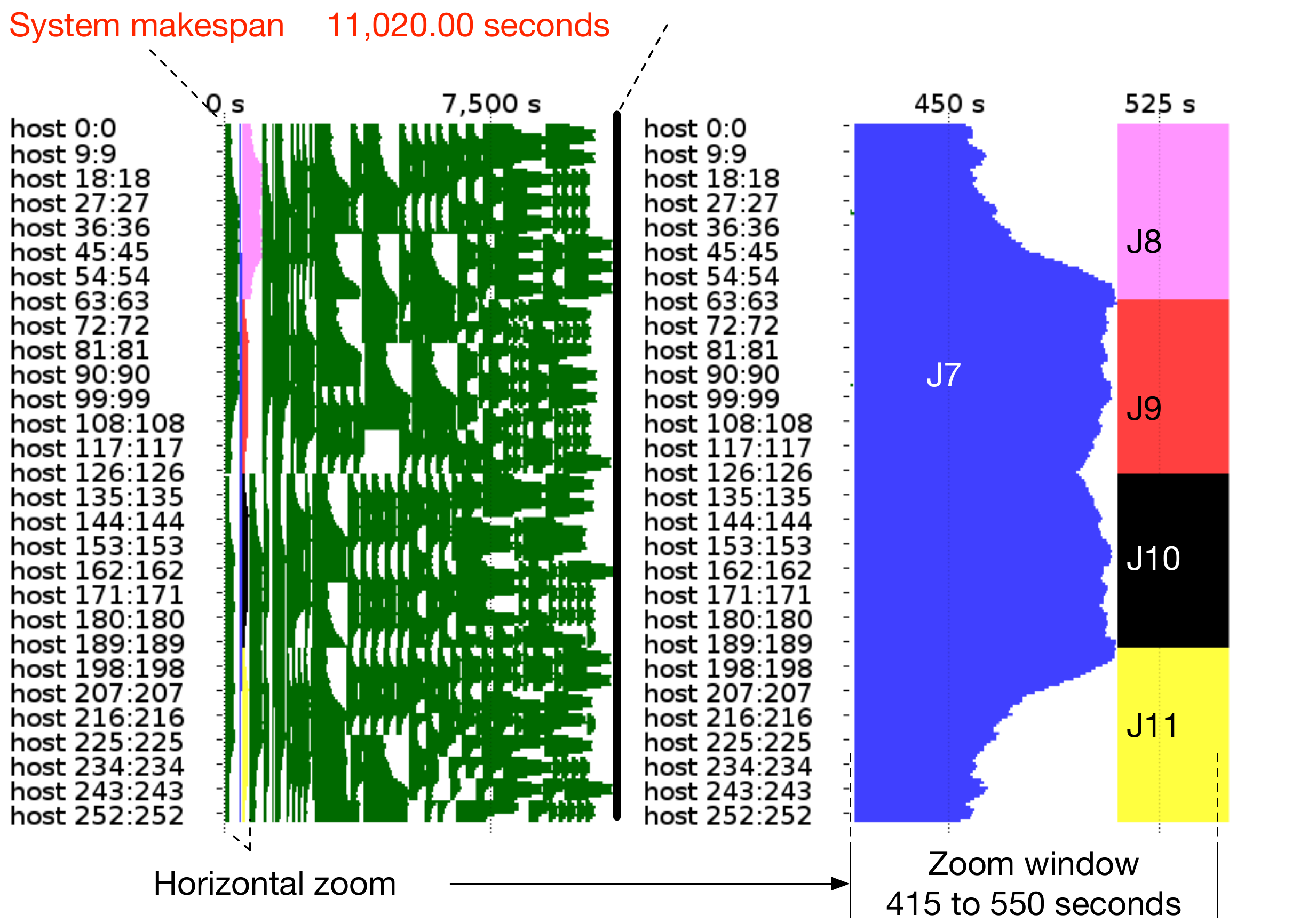}
		\label{fig:woutt}%
	}
	\\
	\centering 
	\subfloat[Improved system utilization with \ourapproach{} at 84.82\%, while J7 \newline approaches completion. J8 and J9 start earlier and utilize \newline the idle resources] {%
		\includegraphics[width=0.8\textwidth, clip,trim=0cm 0cm 0cm 0cm]{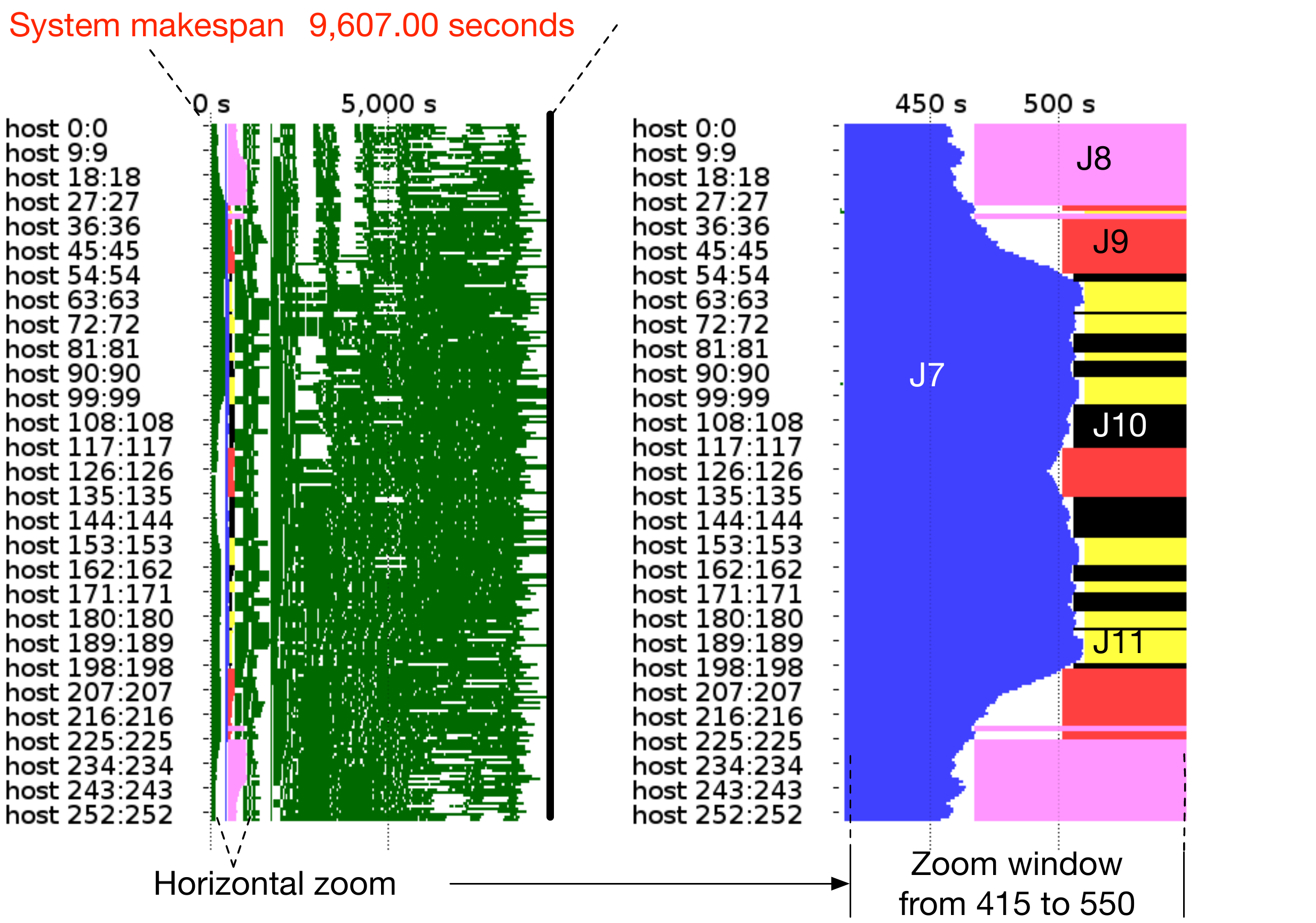}
		\label{fig:withj}%
	} 
	\caption{Visualization (obtained using Vampir~\cite{Vampir}) of the execution trace of the \mbox{ESP-Mandelbrot} workload. STATIC is used at the ALS, while FCFS+BF is used at the BLS. White spaces indicate idle computing resources, while colors denote executing jobs. The short timeline on the right side of each \mbox{sub-figure} is a zoom into a certain time interval of the timeline on the left side. }
\end{figure*}

For \mbox{ESP-Mandelbrot}, Figure~\ref{fig:SU-MLBT} shows that \ourapproach{} increased the average system utilization when the jobs used STATIC from 71.2\% to 83.82\%.
When jobs are executed using GSS and AF, \ourapproach{} is only able to increase the average system utilization by 0.5\% and 0.05\%, respectively. 
This is because AF can achieve a highly balanced execution of all jobs\out{ (see Figure~\ref{fig:ESP-MLBT})}.
By enabling our resourceful scheduling approach, the system makespan of the \mbox{ESP-Mandelbrot} using STATIC is reduced from \num{11020} to 8,840 seconds (\ahm{see} Figures~\ref{fig:SU-MLBT-wout} and~\ref{fig:SU-MLBT-with}).

When all jobs are highly load balanced, our approach offers slight improvements in terms of increasing system utilization.
%For instance, when executing \mbox{ESP-PSIA} and \mbox{ESP-Mandelbrot} jobs using AF, the system utilization   	
However, the slight improvements in system utilization are of high value for HPC operators as they translate into efficient power consumption~\cite{sarood2014maximizing}.
Future work will explore the relation between \ourapproach{} and the power consumption efficiency.

Because of the new feature we added to the Slurm simulator~\cite{simakov2017slurm}, we are also able to visualize the execution trace of the workload at  coarse- and \mbox{fine-grain} scales.
The left side of Figure~\ref{fig:woutt} shows the entire \mbox{ESP-Mandelbrot} execution trace in which STATIC is used at the ALS, FCFS+BF is used at the BLS, and the \emph{proposed resourceful ordination approach is not enabled}. 
The right side of Figure~\ref{fig:woutt} is a horizontal zoom into the timeline of the execution trace between 415-550 seconds.
Zooming this close helps to understand the poor system utilization, i.e., why jobs J8, J9, J10, and J11 wait for the latest computing resources of job J7 to become free.

Figure~\ref{fig:withj} shows the execution trace of the same scenario (STATIC at ALS and FCFS+BS at BLS) with the \emph{proposed resourceful coordination approach enabled}.
At the \mbox{coarse-grain} time scale (left side), the intensity of the green color (busy computing resources) is higher in Figure~\ref{fig:withj} than Figure~\ref{fig:woutt}. 
The total system makespan is shorter in Figure~\ref{fig:withj} than Figure~\ref{fig:woutt} by 1,413 seconds. 
On the right side of Figure~\ref{fig:withj} (horizontal zoom from 415 to 550 seconds), due to the usage of the \emph{proposed resourceful coordination approach}, jobs J8, J9, J10, and J11 started earlier than in Figure~\ref{fig:woutt}.
This reduces the idle times of the computing resource and increases the overall system utilization.

Jobs~J8, J9, J10, and J11 in Figure~\ref{fig:withj} are assigned to \mbox{non-contiguous} hosts compared to their resource allocation in Figure~\ref{fig:woutt}.
In practice, such a \mbox{non-contiguous} resource allocation may cause performance degradation for \mbox{communication-intensive} applications.
The applications PSIA and Mandelbrot used in the current work are \mbox{computationally-intensive}.
Therefore, such a \mbox{non-contiguous} allocation bears no effect on their simulated performance.
Future work will include further analysis of the performance advantages and disadvantages of \ourapproach in the case of mixed types of applications (computation-, communication-, and \mbox{I/O-intensive} applications).

%\begin{figure}[!t]
%	\includegraphics[width=\columnwidth, clip,trim=0cm 0cm 0cm 0cm]{figures/final2_without}
%	\caption{ Visualization of the execution trace of the \mbox{ESP-Mandelbrot} workload  with Vampir~\cite{Vampir}. STATIC was used at the ALS, while FCFS and BF were used at the BLS. \emph{The proposed resourceful scheduling approach was not enabled.} White spaces indicate idle  computing resources, while colors denote executing jobs.
%		The short timeline on the right side is a zoom into a certain time interval of the timeline on the left side. Note the idle resources while J7 nears completion.}
%	\label{fig:woutt}%
%\end{figure}
%\begin{figure}[!t]
%	\includegraphics[width=\columnwidth, clip,trim=0cm 0cm 0cm 0cm]{figures/final2_with}
%	\caption{Visualization of the execution trace of the \mbox{ESP-Mandelbrot} workload  obtained using Vampir~\cite{Vampir}. STATIC was used at the ALS, while FCFS and BF were used at the BLS. \emph{The proposed resourceful scheduling approach was enabled.} White spaces indicate idle computing resources, while colors denote executing jobs.  
%		The short timeline on the right side is a zoom into a certain time interval of the timeline on the left side. Note the decreased amount of idle resources and the start of J8, while  J7 nears completion.}
%	\label{fig:withj}%
%\end{figure}

%% file: conclusion_futurework.tex
% !TEX root =  paper_index.tex
\section{Conclusion and Future Work}
\label{sec:conclusion}	
%\lipsum[2-4]
The present work proposed a resourceful coordination approach (\ourapproach{}) that allows application schedulers to cooperate by involving the batch scheduler. 
We implemented the proposed approach in a \mbox{two-level} simulation using realistic and \mbox{well-known} simulators (a \mbox{Slurm-based} simulator~\cite{simakov2017slurm} and \mbox{a SimGrid-based} simulator~\cite{mohammed2019approach}). 
The effective system performance~(ESP) benchmark was used to assess the proposed approach by instantiating it with the parallel \mbox{spin-image} generation and the Mandelbrot set. 

Our proposed \ourapproach{} increased the entire system utilization by 12.6\% and decreased the system makespan by the same percent when applications suffered from severe load imbalance. 
When application execution was balanced, e.g., when employing AF for ESP-Mandelbrot, \ourapproach{} increased the entire system utilization by 0.5\% as there were few idle system resources to exploit. 
These improvements at the system level are important to eliminate unnecessary system waste, and consequently, unnecessary energy waste, which instead could be used to support small cities~\cite{sarood2014maximizing}.
%
% \mbox{large-scale} HPC system \textbf{wastes} only 1\% to 10\% of its computing cycles, it wastes energy that could support a small city~\cite{sarood2014maximizing}.}
 %
%System utilization was slightly improved by 0.05\% when applications had balanced execution.
%Therefore, the improvements achieved by \ourapproach{} are of high value for HPC operators. % as they translate to efficient power consumption~\cite{sarood2014maximizing}.
%
The present work also shows that for \mbox{long-executing} HPC applications, exploiting computing resources' idle times (in the order of a few seconds) can significantly improve the entire system utilization.
Prior to this work, it was commonly accepted that short computing resource idle times should be filled by Big Data workloads~\cite{mercier2017big}.
The current work highlighted the potential of exploring such idle times also for HPC workloads.

Our extensions to the \mbox{Slurm-simulator}~\cite{simakov2017slurm} enabled the visual analysis of the workload execution at coarse- and \mbox{fine-grain} temporal resolutions using Vampir~\cite{Vampir}. 
The visual analysis showed that using our approach idle resources were exploited efficiently, while jobs were not assigned to contiguous computing resources.
Such a \mbox{non-contiguous} resource allocation may cause performance degradation for \mbox{communication-intensive} applications which were not in the scope of the present work but planned as future work.

Future work also includes porting our changes to the latest source code of Slurm. 
This porting will enable RCA assessment in a real production environment, including more HPC applications. % and to the latest version of the DLS4LB tool~\cite{mohammed2019simas}.
Communication between the application schedulers and the batch system requires standardization. 
Several APIs offer such communication~\cite{d2018drom,prabhakaran2015batch} and will be evaluated in future work.
Since our approach does not depend on a specific RJMS, we plan to explore the integration of \ourapproach{} approach into modern and future RJMS, such as Flux~\cite{FLUX3}.
Extend \ourapproach{} to include information exchange about other computing resources, such as GPUs and co-processors, is also planned for the future.
%Last but not least, we plan to push our extended Slurm simulator forward by modeling the scheduling overhead of the internal Slurm operations.

\section*{Acknowledgments}
This work has been in part supported by the Swiss National Science Foundation in the context of the ``Multi-level Scheduling in Large Scale High Performance Computers'' (MLS) grant, number 169123,  the Swiss Platform for Advanced Scientific Computing (PASC) project ``SPH-EXA: Optimizing Smoothed Particle Hydrodynamics for Exascale Computing'', and by DAPHNE, funded by the European Union's Horizon 2020 research and innovation programme under grant agreement No 957407.
%This work has been partially supported by the Swiss Platform for Advanced Scientific Computing (PASC) project SPH-EXA: Optimizing Smooth Particle Hydrodynamics for Exascale Computing and by the Swiss National Science Foundation in the context of the ``Multi-level Scheduling in Large Scale High Performance Computers" (MLS) grant, number 169123.